# Numerical simulation of lunar response to gravitational waves and its 3D topographic effect using the spectral-element method

Lei Zhang[1], Han Yan[2], Jinhai Zhang[1,*], Xian Chen[2,3]

[1] Center for Deep Earth Technology and Equipment, Key Laboratory of Deep Petroleum Intelligent Exploration and Development, Institute of Geology and Geophysics, Chinese Academy of Sciences, Beijing 100029, China

[2] Department of Astronomy, School of Physics, Peking University, Beijing 100871, China

[3] Kavli Institute for Astronomy and Astrophysics, Peking University, Beijing 100871, China
*Correspondence to: zjh@mail.iggcas.ac.cn

**ORCID**: Lei Zhang: 0000-0002-6067-3325; Han Yan: 0000-0003-1320-5243; Jinhai Zhang: 0000-0001-6314-5299; Xian Chen: 0000-0003-3950-9317.

**ABSTRACT**. The Moon has been regarded as a natural Weber bar capable of amplifying gravitational waves (GWs) for detecting events across a wide range of frequencies. However, accurately determining the amplification effects remains challenging due to the absence of 3D numerical simulation methods. In this study, we develop a high-order 3D finite element method (spectral-element method, SEM) to numerically simulate the lunar response to GWs below 20 mHz. We verify the accuracy of our method by comparing the resonant peaks of our results with those from semi-analytical solutions and find that the frequency deviation is less than 3% for the first peak at about 1 mHz and less than 0.8% for the subsequent peaks up to 10 mHz. Using this method, we evaluate the amplification of GW signals due to 3D topographic effects of the Moon, and we find enhancements at a series of specific frequency components. These results highlight the non-negligible effect of surface topography on the lunar response to GWs, as a fundamental factor that holds significant implications across both global and regional analyses. Our work paves the way for a comprehensive evaluation of the Moon's resonant response to GWs, helpful for the strategic planning of lunar GW detections.

## I. INTRODUCTION

The Moon is thought to have greater potential for amplifying the gravitational-wave (GW) signals in decihertz frequency band, compared with ground- and space-based laser interferometry [1-5], and pulsar timing array [6-9], especially for detecting astrophysical sources caused by supernovae, compact binaries, intermediate-mass black holes, intermediate mass-ratio inspirals, and stochastic GW backgrounds [10-12]. As the ideas of detecting GWs on the Moon become increasingly popular [13-15], strategic pre-mission planning—encompassing site selection, instrument configuration, parameter optimization, and landing zone assessment—has become imperative and must be tightly coupled with specific scientific objectives to ensure mission success. However, it remains unclear where the ideal landing zone is for deploying GW detectors, due to the difficulty in evaluating the actual response to GWs on the Moon.

Previous studies have considered the layered interior and/or subsurface structures of the Moon [16-20], and showed that the Moon has great potential for amplifying the GWs at multiple resonant frequencies [18,20]. However, their approaches are based on ideally spherical models thus cannot include the effect caused by fluctuating topography and crustal thickness of the Moon. To solve this issue, Zhang et al. [21] employed a two-dimensional (2D) FEM to numerically simulate the Moon's response to GWs. Their work points to a new way of accurately simulating the realistic lunar response to GWs in the future. However, since their model is 2D, it could not account for real three-dimensional (3D) amplification effects of both topographic fluctuations and interior lateral heterogeneity of the Moon. Consequently, the resonant peaks from their 2D simulations apparently deviate from the 3D semi-analytical results [21]. Therefore, it is necessary to develop 3D FEM to clarify the actual lunar response to GWs.

Here, we construct a 3D lunar FEM model and develop a high-order FEM (spectral-element method, SEM) to numerically simulate the Moon's response to GWs. We also compare our numerical results with semi-analytical solutions (Section II) to verify the accuracy of the proposed method. Then, we evaluate 3D topographic effects by comparative analyses of the lunar response to GWs with and without the incorporation of realistic lunar topography (Fig. 1, Section III). In addition, we discuss the limitations of our current method and point out potential directions for future development (Section IV). Finally, we draw a conclusion to this work (Section V).

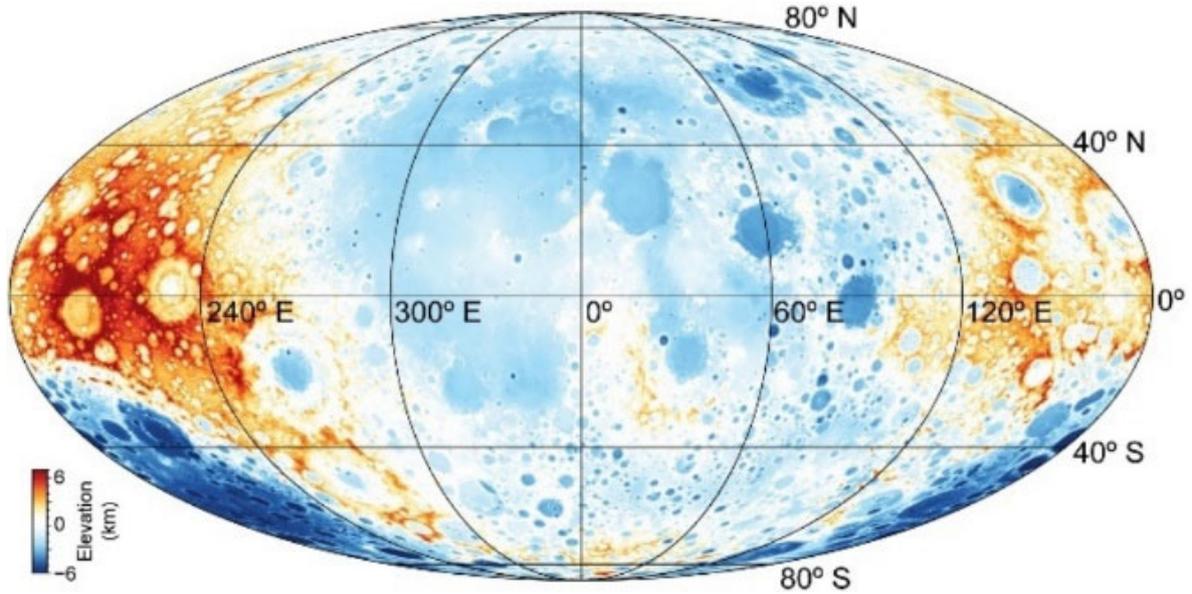

**FIG. 1. Lunar topography model.** The elevation data is derived from Lunar Orbiter Laser Altimeter (LOLA, https://pds-geosciences.wustl.edu/missions/lro/lola.htm).

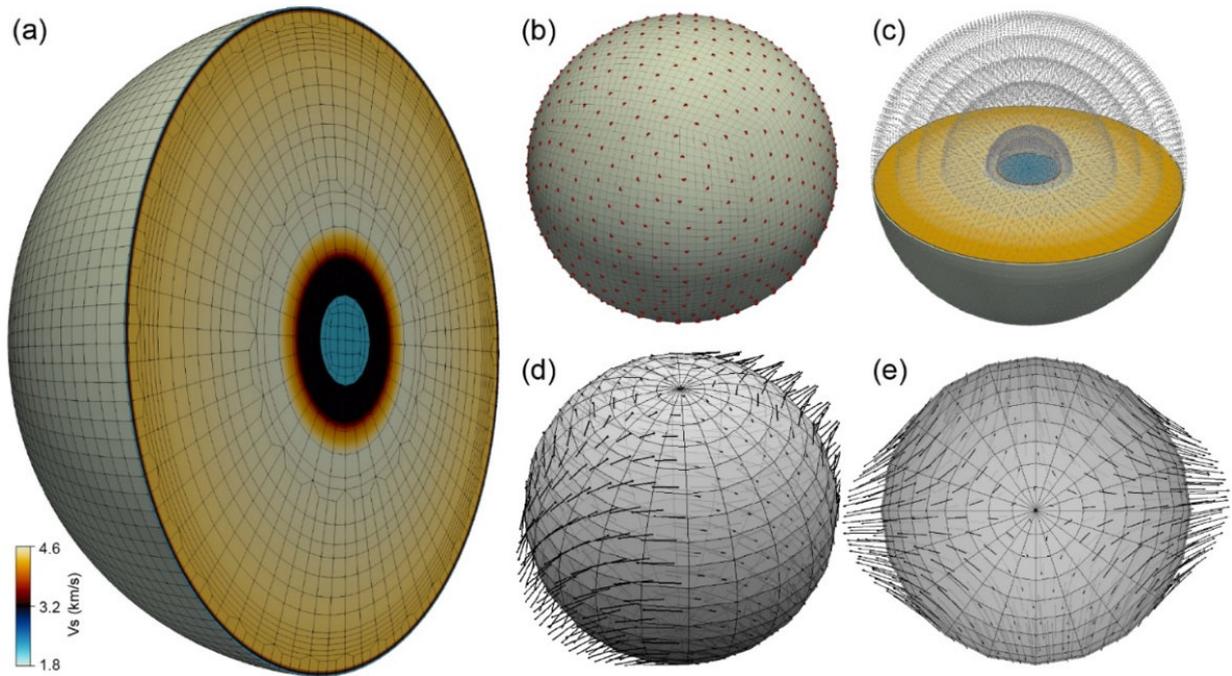

**FIG. 2. The spectral-element model for the entire Moon and the distribution of the Dyson-type GW sources.** (a) The model used for GW simulation. The cross section of the global model shows the interior structure of the Moon. (b) The distribution of the receivers (red dots) on the surface of the Moon. (c) The distribution of the sources induced by GW (gray dots). (d) The force density imposed on the surface and the internal interfaces of the Moon. Only 500 vectors are plotted here for a better display. (e) The same as (d) but with a different viewing angle.

## II. METHODS AND BENCHMARKS

### 2.1 Method

The theory of calculating the force density imposed by GW on an elastic body was first laid down by Dyson [22], who introduced a coupling term between GW and elastic body and derived an external-force density,

$$\vec{f} = \nabla \cdot (\mu \mathbf{h}), \tag{1}$$

where $\mu$ is the shear modulus and $\mathbf{h}$ refers to 3D spatial components of the GW tensor. Based on the previous studies on analytical and semi-analytical solutions of planet's response to GWs [18, 23, 24], the GW-induced elastic force density acting on the Moon can be expressed as

$$\mathbf{f}(\mathbf{x}, t) = h_0 \frac{\partial \mu}{\partial r} e^{i\omega_g t} \boldsymbol{\epsilon} \cdot \hat{\mathbf{r}}, \tag{2}$$

with

$$\hat{\mathbf{r}} = \left( \frac{x}{\sqrt{x^2+y^2+z^2}}, \frac{y}{\sqrt{x^2+y^2+z^2}}, \frac{z}{\sqrt{x^2+y^2+z^2}} \right), \tag{3}$$

in which the polarization tensor $\epsilon$ can be further expressed as

$$\boldsymbol{\epsilon} = (\mathbf{l} \otimes \mathbf{l} - \mathbf{m} \otimes \mathbf{m}) + (\mathbf{l} \otimes \mathbf{m} + \mathbf{m} \otimes \mathbf{l}), \tag{4}$$
$$\mathbf{l} = \cos \nu (-\hat{\mathbf{e}}_e) + \sin \nu (-\hat{\mathbf{e}}_\lambda), \tag{5}$$
$$\mathbf{m} = \sin \nu (\hat{\mathbf{e}}_e) + \cos \nu (-\hat{\mathbf{e}}_\lambda), \tag{6}$$

where

$$\hat{\mathbf{e}}_e = (\cos e \cos \lambda, \cos e \sin \lambda, -\sin e)$$
$$\hat{\mathbf{e}}_\lambda = (-\sin \lambda, \cos \lambda, 0) \tag{7}$$
$$\hat{\mathbf{e}}_k = (\sin e \cos \lambda, \sin e \sin \lambda, \cos e),$$

are the orthogonal unit base vectors for GW propagating along $\hat{\mathbf{e}}_k$.

Based on a spherically layered model [18,21], we add the force densities (i.e., surface force density and body force density) at the interface where the shear modulus and S-wave velocity vary significantly. We also add force vectors at the lunar surface in accordance with the discontinuous boundary conditions.

### 2.2 Numerical model

Here we describe the 3D SEM which we use to simulate the lunar response to GWs. Fig. 2 presents the spectral-element model constructed for the entire Moon and illustrates the key parameters of the GW simulation setup. Our numerical model and simulation are based on ABAQUS [25] and SPECFEM3D codes [26-28] and we incorporate GW-generated forces into the simulation. A global 3D lunar model with 44,512 fourth-order spectral elements has been constructed [Fig. 2(a)]. This model is adapted from references [18, 21]. Up to 800 receivers are uniformly set on the lunar surface for recording the lunar response to GWs.

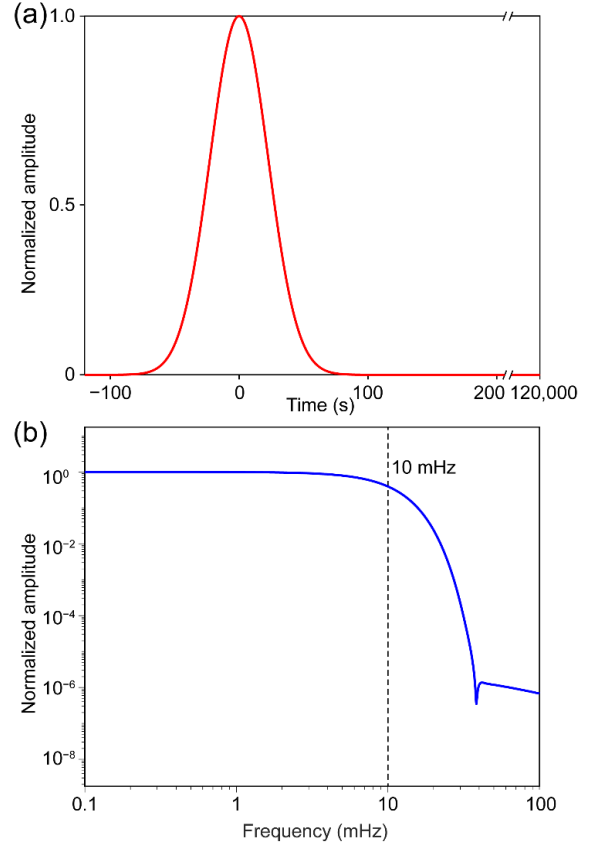

**FIG. 3. Source time function (dimensionless) with a cutoff frequency at 10 mHz.** (a) Source time function with a duration of 120,000 s. (b) The normalized spectrum.

To minimize numerical dispersion artifacts that may arise from coarse discretization, the grid size is set to approximately 70 km in the horizontal direction at the surface and increases proportionally with seismic wave velocities [Fig. 2(a)]. A series of numerical experiments (not shown here) show that the current model can accurately resolve frequencies up to 20 mHz. A source time function (STF) is used to excite lunar response (Fig. 3). To cover a broad frequency range [21], the Gaussian wavelet STF with a half-duration of 50 s is adopted [21]. The total simulation time duration is up to 120,000 s (Fig. 3), enabling seismic waves to propagate multiple times through the entire Moon volume. From the perspective of numerical stability, the maximum allowable time step is approximately 0.2 s. The 3D numerical simulation was conducted on 256 cores in parallel mode at the National Supercomputing Center in Wuxi, China, whose wall-clock time is about 1.5 hours.

**2.3 Results**

Fig. 4 shows the displacement amplitudes of the lunar surface in response to GWs at several representative time steps, in both 3D and 2D visualizations. The seismic waves induced by GWs of a given polarization can propagate throughout the entire Moon, resulting in global-scale oscillations. This process excites the Moon's normal modes, which can be observed on the lunar surface. The GW-induced responses also exhibit significant temporal variations on the lunar surface; in addition, a clear and persistent quadrupolar feature remains observable, especially at $t$ = 10,000 s, 20,000 s and 30,000 s ([Fig. 4(a~c)]). This feature agrees well with theoretical predictions [29], thereby validating the feasibility of our numerical model.

Fig. 5 compares our numerical lunar response function with the previous semi-analytical solutions based on MINEOS calculations [18]. Our simulated displacement responses from 800 uniformly distributed receivers on the entire lunar surface were individually Fourier-transformed. Then, after direct current shift removal, the Fourier spectra were averaged across all receivers to produce a 'mean' spectral response for three component directions. The rationale for this simple averaging method lies in the theoretical expectation that, on a spherically symmetric Moon, the response at each station is modified only by a correction coefficient dependent on polarization-angle [18]. This correction remains frequency-independent under spherical symmetry.

Our numerical results demonstrate excellent agreement with the semi-analytical solution at low frequencies (below 10 mHz). The spectral peaks at about 1, 2, 3, and 4 mHz closely match the results obtained by the semi-analytical solution (Fig. 5), both in their frequency positions and spectrum shapes. In particular, the frequency difference is less than 3% for the first peak (1 mHz) and under 0.8% for the subsequent 9 peaks (Table 1). While the peaks match, some differences remain in the spectral troughs, suggesting limitations in the dynamical range of the simulation results, probably due to the coarse grid size of 70 km.

**III. EVALUATION OF THE 3D LUNAR TOPOGRAPHIC EFFECT**

After validating the model and methodology (Figs. 2-5), in this section, we investigate the influence of 3D topographic effects on the lunar response to GWs. We construct a 3D topographic model of the Moon [30] by radially adjusting the positions of the surface nodes in the established finite-element mesh (Fig. 2). The locations of the applied surface force density and the receivers were updated accordingly to maintain their correct positions relative to the new topography. Fig. 6 compares the GW-induced surface displacement amplitudes with and without topography at several representative time steps. In the early propagation stage (e.g., 8000 s) [Fig. 6(a)], on a large scale, the general pattern of the response to GWs remains unchanged when surface topography is introduced, though local fluctuation of amplification factor can be seen. At later stages, significant differences in the amplification pattern emerge and intensify over time [31-33]. In several instances (e.g., 116,000 s) [Fig. 6(d)], an amplification factor exceeding 2 is observed across most of the lunar surface, indicating that given a polarization angle, topographic effects vary with the increase of propagation time.

**Table 1. Comparison of our numerical and semi-analytical resonant peaks**

| # | Semi-analytical (MINEOS) results /mHz | Numerical results /mHz | Error |
|---|---|---|---|
| 1 | 1.09144 | 1.06667 | −2.269% |
| 2 | 1.92309 | 1.90833 | −0.768% |
| 3 | 3.13329 | 3.13333 | 0.001% |
| 4 | 4.30527 | 4.29167 | −0.316% |
| 5 | 5.34564 | 5.33333 | −0.230% |
| 6 | 6.33870 | 6.35833 | 0.310% |
| 7 | 7.41310 | 7.36667 | −0.626% |
| 8 | 8.43335 | 8.36667 | −0.791% |
| 9 | 8.79023 | 8.78333 | −0.078% |
| 10 | 9.59401 | 9.57500 | −0.198% |

Fig. 7 shows the topographic amplification effects near the frequencies of resonant peaks (specifically the first four modes at 1, 2, 3, and 4 mHz) derived from the responses of 800 globally distributed receivers on the lunar surface. As we can see, in the absence of topographic variations ('flat', left panels, Fig. 7), the modal patterns exhibit high symmetry. When topography is introduced, these patterns become perturbed ('topo', middle panels, Fig. 7). To quantify the amplification effect, the topographic modal responses were divided by the non-topographic (flat-surface) ones, providing amplification factors around the resonant peaks. These factors are generally greater than 1, indicating a consistent effect of amplification, although the excess remains within 10% (Fig. 7). This result indicates that topography generally favors the amplification of lunar responses. Furthermore, the amplification effect appears on a hemispheric scale. The lack of small-scale amplification may be attributed to the lack of small-scale topographic features in the current model. Note that Fig. 7 displays displacement amplitudes from the 800 chosen receivers. Therefore, the spatial resolution is substantially lower than that of Fig. 6 which uses all nodes on the entire lunar surface for visualization.

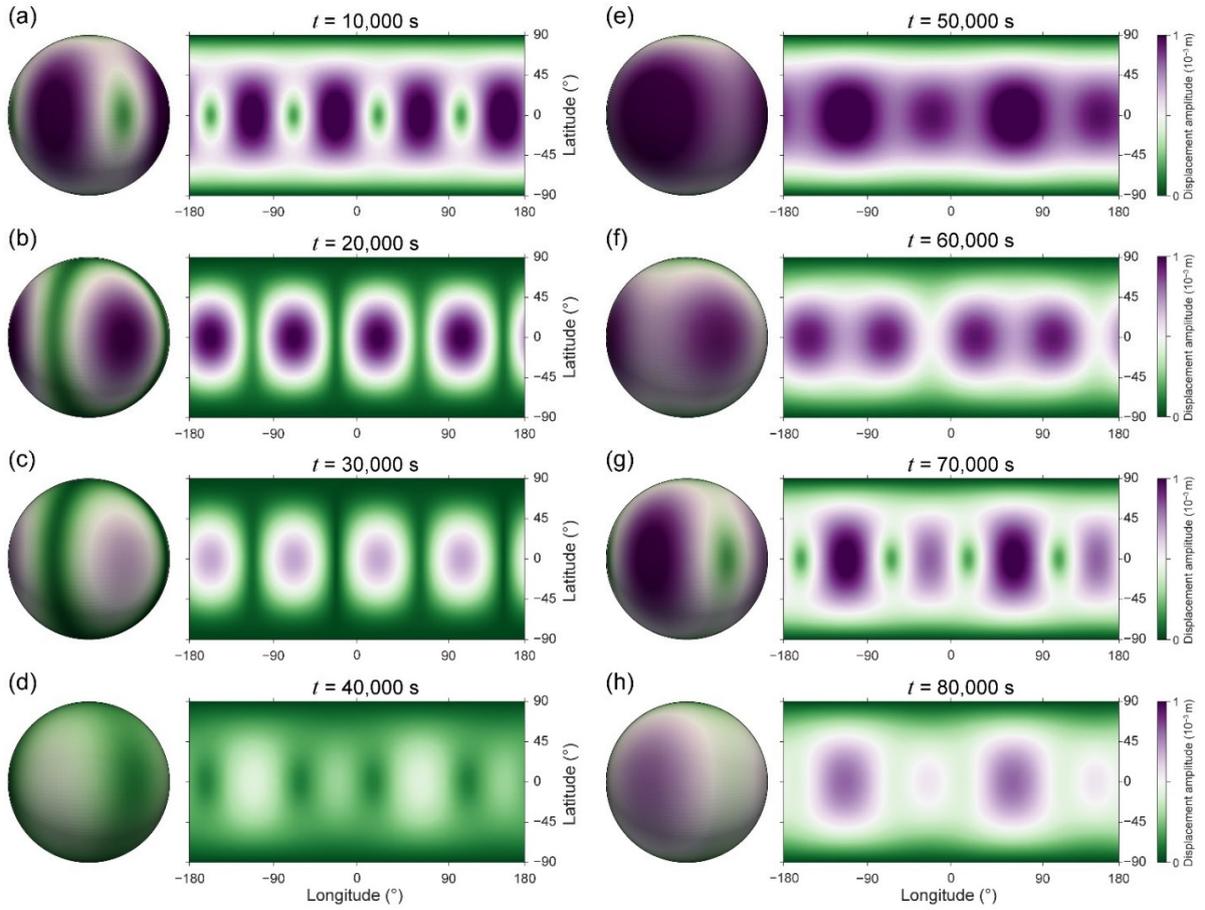

**FIG. 4. Representative snapshots of the displacement amplitude on lunar surface induced by passing GWs.** (a) $t$ = 10,000 s; (b) $t$ = 20,000 s; (c) $t$ = 30,000 s; (d) $t$ = 40,000 s; (e) $t$ = 50,000 s; (f) $t$ = 60,000 s; (g) $t$ = 70,000 s; (h) $t$ = 80,000 s. Both 3D (left) and 2D (right) visualizations are displayed.

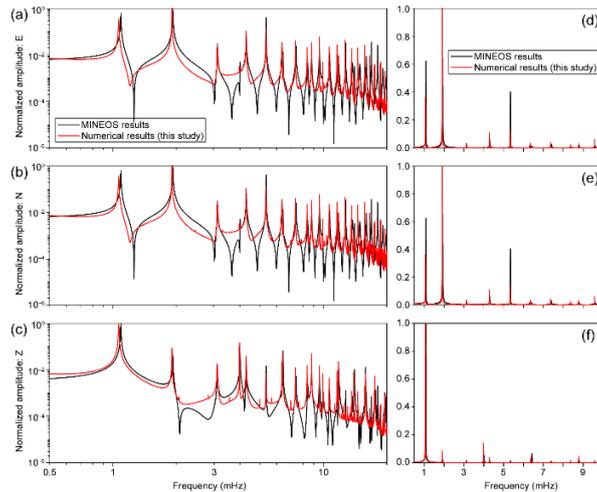

**FIG. 5. Response function from our FEM simulations (red) and the MINEOS semi-analytical calculation (black).** (a) E-W direction component; (b) N-S direction component; (c) Z direction component. Notice that the results are normalized by the maximum values of the response functions. (d~f) are the same as (a~c) but in normal (instead of log) axis.

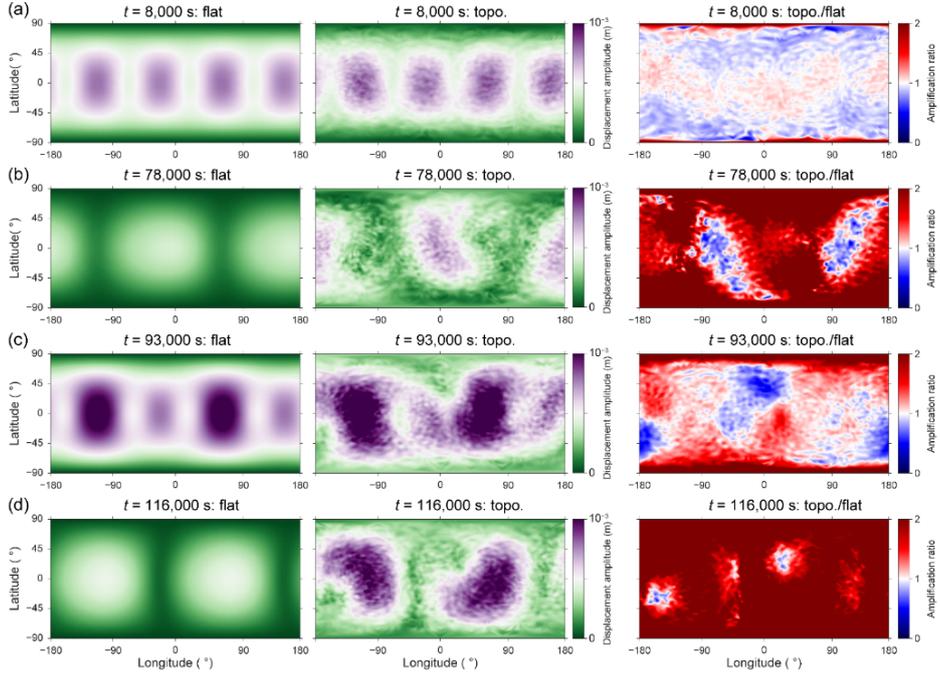

**FIG. 6. 3D Topographic effects on the lunar response to GWs.** The left panels show the displacement amplitudes without lunar topography (flat); the middle panels are displacement amplitudes with lunar topography (topo); and the right panels are the amplification factors (ratios of these two results) in the presence of the topographic fluctuation. (a) $t = 8000$ s; (b) $t = 78,000$ s; (c) $t = 93,000$ s; (d) $t = 116,000$ s.

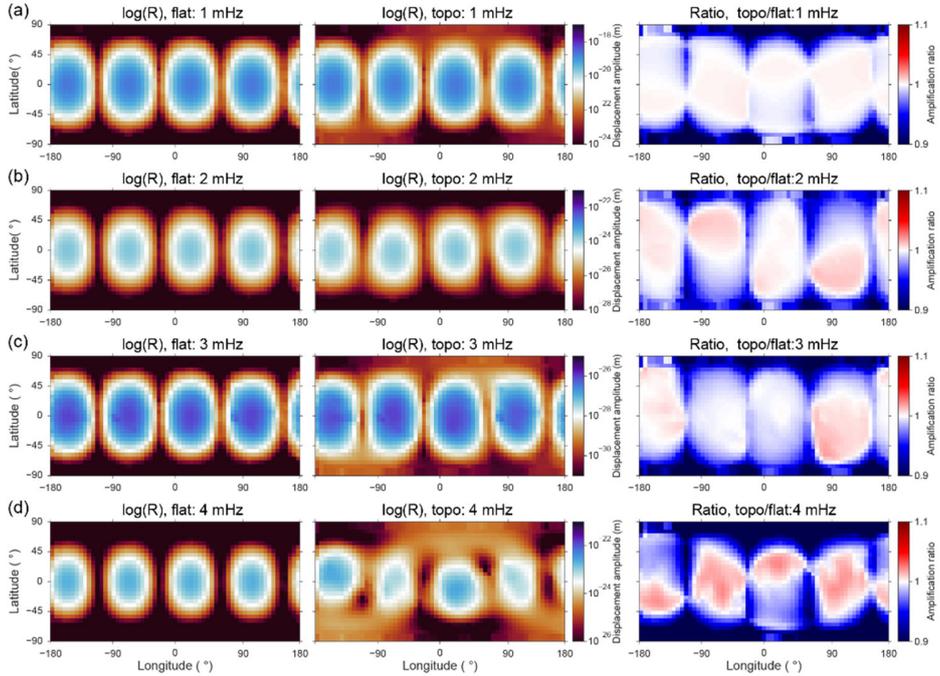

**FIG. 7. Frequency-dependent topographic effects at resonant frequencies of lunar response to GWs.** The left panels are the displacement amplitude without lunar topography (flat); the middle panels are vertical displacements with lunar topography; the right panels are the amplification ratios caused by the topographic effect (topo). From top to bottom, the corresponding frequencies are centered at (a) 1 mHz; (b) 2 mHz; (c) 3 mHz; (d) 4 mHz, respectively. These four frequencies are related to the peak frequencies of the response functions shown in Fig. 5.

To further identify the frequency ranges affected by topographic effects at a global scale, we generated a 'mean' spectral response for the topographic model (Fig. 8) following the same signal processing methods as in Fig. 5. Comparative analysis shows that in the topographic model, spectral amplitudes display minor, isolated enhancements at certain frequencies (Fig. 8). However, these newly emerged minor peaks do not align with the original resonant peaks. Moreover, broadband amplification is evident in several spectral troughs, for instance, around 1.2 mHz (E-W and N-S components), 2–3 mHz (N-component), and 3–5 mHz (Z-component). The topographic amplification effect grows notably near 10 mHz, suggesting a stronger influence of topography on higher-frequency responses, which aligns with theoretical expectations.

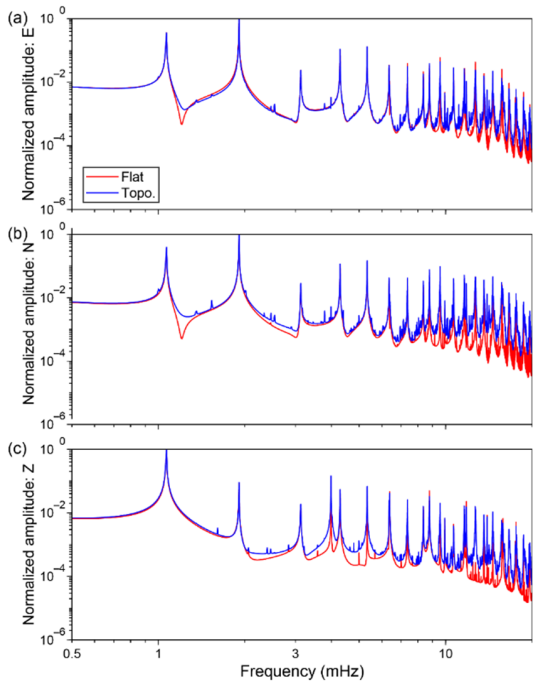

**FIG. 8. Modification of lunar response function by topography.** (a) E-W direction component; (b) N-S direction component; (c) Z-direction component. The response functions are all normalized.

Fig. 9 shows the topographic amplification effects across the southern hemisphere, centered at the South Pole, a region of high scientific interest due to its geological background and planned deployment of future seismometers (e.g., Chang'e-7, Farside Seismic Suite and Artemis III) [34-36]. A comparison between the models with and without topography reveals distinct spatiotemporal patterns of amplification. At the early propagation stage (e.g., $t = 20,000$ s), the amplification factor exhibits a cross-like distribution, primarily resulting from the minimal response in the flat model. By the intermediate stage (e.g., $t = 64,000$ s), significant amplification factors (>1) appear mainly near the South Pole and within the South Pole–Aitken (SPA) Basin, with moderate amplification observed in the Apollo Basin. In contrast, an amplification factor below 1 is observed in the regions such as the Mare Nectaris, Schrödinger, and Planck basins. In the late stage (e.g., $t = 100,000$ s), amplification becomes widespread across nearly the entire southern hemisphere. The amplification factors exceed 2 in prominent areas, including the South Pole–Aitken (SPA) Basin, Schrödinger Basin, Planck Basin, and Mare Fecunditatis, indicating their strong potential for topographic amplification given a certain polarization angle.

## IV. DISCUSSIONS

In our current simulation, we investigate the topographic amplification effects on lunar seismic responses induced by GW sources with a specific polarization angle. The amplification factors reported here are derived under this specific polarization angle; thus, artifacts in the amplification factor could arise in the regions where the model without topography exhibits minimal response. This emphasizes the limitations of a single-polarization-angle approach in quantifying the true topographic effect. To accurately quantify the lunar topography on its seismic response to GWs, multiple polarization angles of the GW sources should be considered in future works.

Our results demonstrate the feasibility of simulating lunar response to GWs using 3D SEM. Due to the current computational constraints, a relatively coarse mesh resolution (~70 km at the lunar surface) was adopted, limiting the simulated frequency range to approximately 20 mHz. While this resolution suffices for validating low-frequency global responses and large-scale topographic effects, it restricts the analysis of higher-frequency responses. Future works should prioritize developing finer-scale meshes or multi-scale numerical simulation methods [37,38] capable of resolving decihertz frequencies, which are critical for a more thorough assessment of the complementarity of lunar GW detection relative to other methods.

While this paper primarily focuses on simulating topographic effects, it is important to note that the lunar interior exhibits substantial heterogeneity due to its complex volcanic history, geological evolution, and subsequent space weathering processes [39-45]. The proposed methodology holds considerable potential for future investigations into how lateral crustal variations and core-mantle structural asymmetries influence the lunar response to GWs. Incorporating more realistic interior models is expected to bring more refined constraints, thereby providing more comprehensive guidance for the design and operation of future lunar GW detectors, as well as for the selection of landing sites.

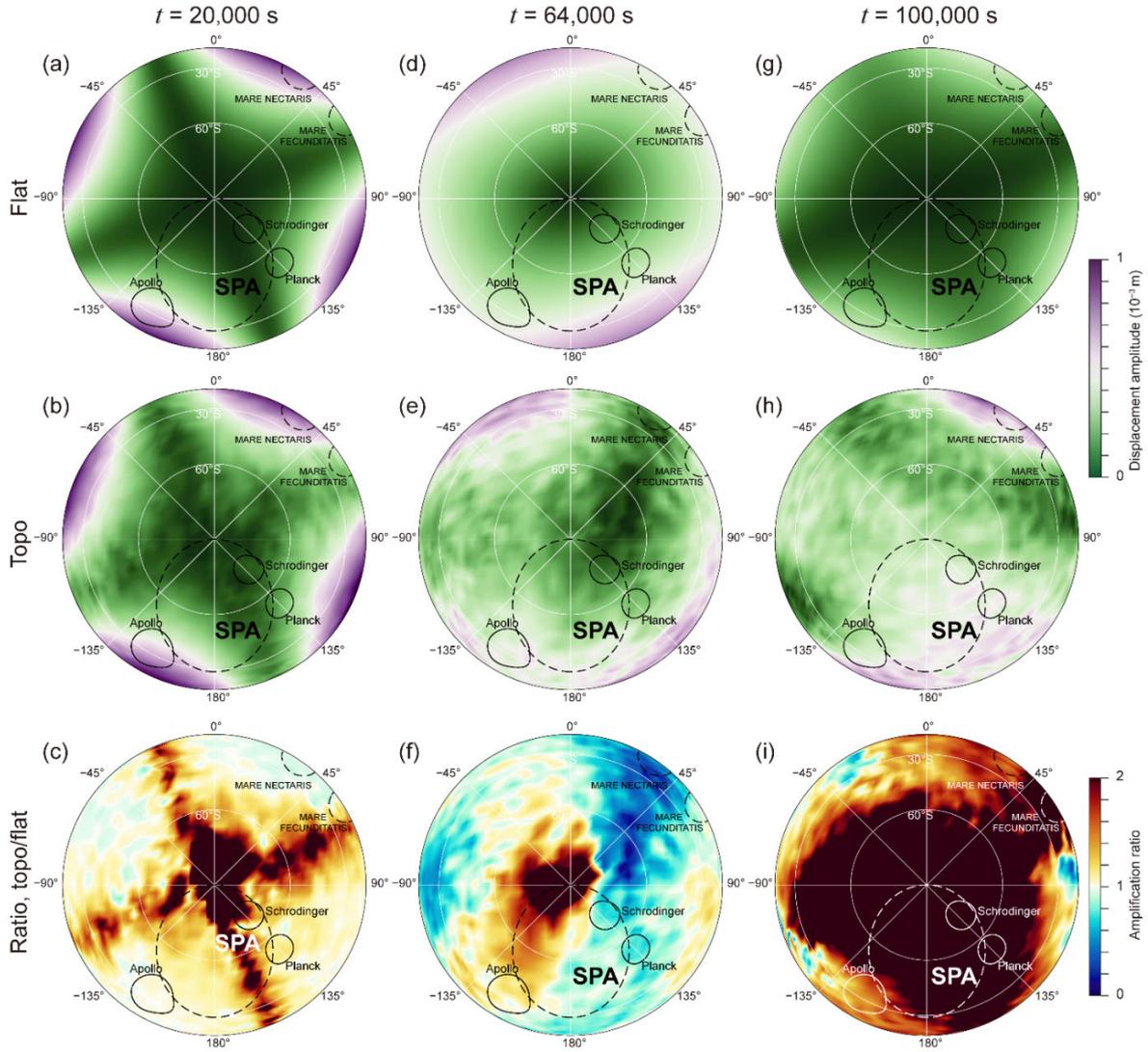

**FIG. 9. Regional topographic effects at the south pole of the Moon.** (a) The displacement amplitude without lunar topography (flat) at $t$ = 20,000 s; (b) The displacement amplitude with lunar topography (topo) at $t$ = 20,000 s; (c) The amplification ratios caused by topographic effects at $t$ = 20,000 s. (d~f) The same as a-c but for at $t$ = 64,000 s. (g~i) The same as a-c but for the results at $t$ = 100,000 s.

## V. CONCLUSIONS

This study presents the first purely numerical 3D simulation of the lunar response to GWs using a high-order FEM—SEM. The method has been validated against semi-analytical solutions in spherically symmetric models, demonstrating its potential for modeling and simulating heterogeneous structures of the Moon. Using this method, we evaluate the amplification of GW signals due to 3D topographic effects of the Moon, which exhibit enhancements at a series of specific frequencies. These results suggest that including the 3D topography of the Moon is critical for the evaluation of the lunar response to GWs as well as for the strategic plan of future lunar GW detections.

## ACKNOWLEDGMENTS

This work is supported by the National Natural Science Foundation of China (Grants No. 42204178, and 12473037), the Key Research Program of the CAS (Grants No. ZDBS-SSW-TLC00104 and KGFZD-145-23-15-2), and the Key Research Program of the Institute of Geology and Geophysics, Chinese Academy of Sciences (Grants No. IGGCAS- 202203, 202401). X.C. acknowledges the support by the National Key Research and Development Program of China (grant No. 2024YFC2207300). H.Y. acknowledges the support by China Scholarship Council (No. 202506010256).

## DATA AVAILABILITY

No data was created or analyzed in this study.